\documentstyle[aps,psfig]{revtex}
\begin{document}
\twocolumn[\hsize\textwidth\columnwidth\hsize\csname
@twocolumnfalse\endcsname
\draft

\title{Coherent quantum oscillations in coupled traps with ultracold atoms}
\author{N.M. Bogoliubov${}^\star$, J. Timonen${}^\dagger$, and
M.B. Zvonarev${}^\star$}

\address{${}^\star$~Steklov Istitute of Mathematics at St.~Petersburg, Russia}

\address{${}^\dagger$~Department of Physics,
University of Jyv\"askyl\"a, Finland}

\maketitle
\begin{abstract}
The dynamics of two traps with ultracold atoms and connected 
by Josephson type coupling, is shown to exhibit a transition from dispersive
dynamics to localized coherent oscillations. This transition is controlled 
by coupling strength and energy offset between the traps. The dynamics is also
shown to be exactly that of a Heisenberg chain with a linear magnetic field.
Possible applications for ``quantum-state engineering'' are discussed. 
 
\end{abstract}
]

Trapping and cooling of atomic gases has produced an interesting laboratory
for analyzing the transition from classical to quantum behavior. The
fascinating new phenomena include the creation \cite{da}
 of Bose-Einstein condensates
in magneto-optical traps, and the realization of coupled quantum systems
with BEC's in  \cite{zol,choi}, e.g., optical lattices.
 One can now analyze in some detail the properties of
Josephson currents  \cite{choi,cat}
and Bloch oscillations \cite{buch} that appear in the coupled
systems, and entanglement of quantum states in individual traps. Possible
use \cite{ci} of trapped ultracold atoms in realizing a quantum computer has also been
considered \cite{lu} (for ``quantum-state engineering'' with
conventional Josephson junctions see, e.g., the recent review \cite{mak}).

It is evident that quantum dynamics of these systems has become \cite{z} an 
important topic.
This is particularly so if one tries to manipulate the dynamics as is
needed, e.g., in the case of quantum computer. Also, rapid decoherence of
oscillatory dynamics even without external perturbations would make
controlled operations very difficult to perform. In coupled atom-field
systems, such as, e.g., the one of two-level atoms interacting with a high-Q
cavity, escaping photons form \cite{rai} the main loss mechanism for quantum
coherence independent of the internal dynamics of the system.
It may thus be useful to consider systems in which controllable quantum dynamics
can be realized without active involvement of a quantized electromagnetic field.

We consider in this Letter a coupled system of two traps with ultracold
atoms, possibly in a condensed state so that they can each be described
with one
boson field $b_{1,2},b^{\dagger}_{1,2}$, with subindeces denoting the trap.
Experiments are now approaching \cite{z} the limit in which a fully quantum
mechanical treatment of the problem is necessary, and it is actually this
limit which is interesting from the standpoint explained above. First
attempts have already been made \cite{ang,pit} to analyze this system quantum mechanically,
and we shall concentrate here on the quantum dynamics of the system, which
so far has remained mostly unexplored. The two traps are assumed to be
connected by a Josephson type of coupling, and atoms are assumed to interact
via delta function potentials so that their interactions are \cite{zol,imam,mil} of
the form $cb^{\dagger}_jb^{\dagger}_jb_jb_j$; $j=1,2$. Here $c$ is the
interaction strength which for analytical simplicity is taken to be of opposite
sign for the two traps (this is now experimentally possible, and results are
expected to be similar for general $c$ albeit much more difficult to achieve
analytically).
 We have previously shown \cite{bt} that for large Josephson coupling a natural quantum
 Hamiltonian for systems of this kind can be expressed in terms of the number
 operators
      $ \hat N_j=b_j^{\dagger }b_j $
 and the exponential phase operators
   $
      \phi _j=(\hat N_j+1)^{-1/2}b_j,\phi _j^{\dagger } = 
      b_j^{\dagger }(\hat N_j+1)^{-1/2}
   $.
 Notice that these phase operators are one-sided unitary:
   $ \phi _j\phi _j^{\dagger }=1 $,
   $ \phi _j^{\dagger }\phi _j=1-|0\rangle_j \langle 0|_j $,
 with $ |0\rangle_j $ appropriate vacua.
 In this case of two coupled traps the Hamiltonian thus becomes 
  \begin{eqnarray}
     \hat H_{tr} &=&c \hat N_1^2-c \hat N_2^2  \label{phg} \\
     &&+(\omega +\tilde \Delta )\hat N_1+\omega \hat N_2-\frac g2\left( \phi
     _1\phi _2^{\dagger }+\phi _1^{\dagger }\phi _2\right) ,  \nonumber
   \end{eqnarray}
    in which
      $ \omega $ and $ \omega +\tilde\Delta $
 are the single particle energies (frequencies) in the two traps.
 We include an offset $ \tilde\Delta $ in the energies (``detuning'')
 because this is an important parameter in the problem
 as will become evident below. $g$ is the strength of Josephson coupling.
 The total number of particles is conserved,
    $[\hat H_{tr},\hat N_1+\hat N_2]=0$,  
 so we can consider without loss of generality the Hamiltonian
      $\hat H_{tr}-\omega (\hat N_1+\hat N_2)+c(\hat N_1+\hat N_2)^2$,  
   \begin{equation}
      \hat H \equiv
      \Delta \hat N_1-\frac g2 \left( \phi _1\phi _2^{\dagger } +
      \phi_1^{\dagger }\phi _2\right) ,
   \label{ph}
   \end{equation}
 where $\Delta =\tilde \Delta + 2c (\hat N_1 + \hat N_2)$.
 The ordinary (orthonormal) Fock states of the system can be formed from the
 vacuum states by operating with the phase operators, 
   \begin{equation}
      |N_1,N_2\rangle =
      (\phi _1^{\dagger })^{N_1}(\phi _2^{\dagger})^{N_2}
      |0\rangle _1|0\rangle _2
   \label{Fock}
   \end{equation}
 such that
   $
      (\hat N_1+\hat N_2)|N_1,N_2 \rangle =
      (N_1+N_2)|N_1,N_2 \rangle \equiv N|N_1,N_2 \rangle 
   $.
In problems like entanglement of quantum states or quantum computing one is
interested in the dynamics of a state which is not an eigenstate of the
Hamiltonian. The Fock states are not eigenstates of Hamiltonian Eq.~(\ref{ph}), and in view of possible applications of the system, we
consider next the dynamics of Fock states. We find it instructive to
consider the dynamics in terms of the correlator 
 \begin{equation}
      \Phi _{jl}^N(t)=
      \langle j,N-j|e^{-it\hat H}|l,N-l\rangle ,
   \label{es2}
   \end{equation}
 which describes the time evolution of the occupation of particles in the
 two traps, and has an obvious connection to probability of observing a given
 state.

 This correlator can be solved exactly for the Hamiltonian (\ref{ph}). The
 most convenient way to find the solution is to first construct an ``equation
 of motion'' for the correlator, which for fixed subindex $j$ takes the form 
   \begin{equation}
      -i\frac 2g\frac d{dt}\Phi _{jl}^N(t) =
      \Phi _{jl+1}^N(t)+\Phi_{jl-1}^N(t)-\frac 2g\Delta l\Phi _{jl}^N(t) ,
   \label{phe}
   \end{equation}
and a similar equation is found for fixed $l$. The initial condition for the
correlator is $\Phi _{jl}^N(0)=\delta _{jl}$, and the boundary conditions
are $\Phi _{jl}^N(t)=0$ for $j,l=-1,N+1$. Notice that this differential
equation is a generalization of the equation of motion for a lattice particle 
in a linear (electric) field \cite{g}.

Consider first the case of vanishing ``field'' or ``effective detuning'', i.e.
 $ \Delta =0$. In this case Eq.~(\ref{phe}) can easily be solved and we find 
   \begin{equation}
      \Phi _{jl}^N(t)=
      \sum_{k=1}^{N+1}\frac{2e^{-itE_k}}{N+2}
      \sin \left[ \frac{\pi k(j+1)}{N+2}\right]
      \sin \left[ \frac{\pi k(l+1)}{N+2}\right] ,
   \label{hc}
   \end{equation}
in which the (Bloch) spectrum $E_k$ is given by 
$E_k=-g\cos (\pi k/(N+2))$. The time
evolution of any initial state can now be followed. We show in Fig.~1 the
probability $|\Phi _{jl}^N(t)|^2$ for the initial state with equal number of
particles in both traps, $j=l=N/2$.
\unitlength=1mm
\begin{figure}[hbt]
\begin{center}
    \begin{picture}(70,79)
        \put(12,10){\psfig{file=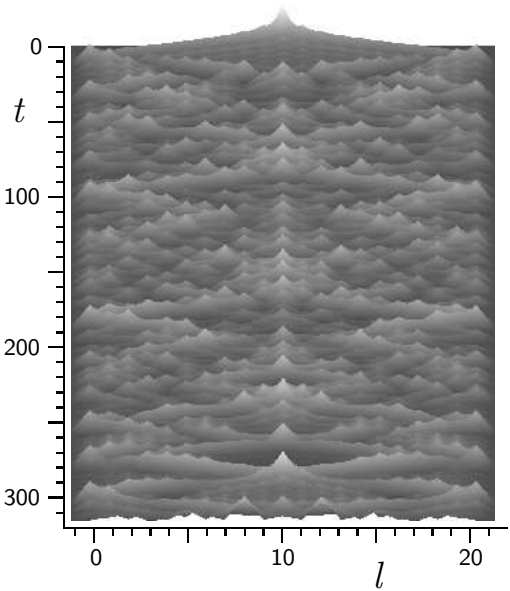,height=68.5mm,width=56mm}}
        \put(11,9){\line(1,0){59}}   
        \put(12.5,9){\line(0,-1){1}}
        \put(14.5,4){\small\sf0}
        \put(15,9){\line(0,-1){2}}
        \put(17.5,9){\line(0,-1){1}}
        \put(20,9){\line(0,-1){1}}
        \put(22.5,9){\line(0,-1){1}}
        \put(25,9){\line(0,-1){1}}
        \put(27.5,9){\line(0,-1){2}}
        \put(30,9){\line(0,-1){1}}
        \put(32.5,9){\line(0,-1){1}}
        \put(35,9){\line(0,-1){1}}
        \put(37.5,9){\line(0,-1){1}}
        \put(38.5,4){\small\sf10}
        \put(40,9){\line(0,-1){2}}
        \put(42.5,9){\line(0,-1){1}}
        \put(45,9){\line(0,-1){1}}
        \put(47.5,9){\line(0,-1){1}}
        \put(50,9){\line(0,-1){1}}
        \put(52.5,9){\line(0,-1){2}}
        \put(55,9){\line(0,-1){1}}
        \put(57.5,9){\line(0,-1){1}}
        \put(60,9){\line(0,-1){1}}
        \put(62.5,9){\line(0,-1){1}}
        \put(63.5,4){\small\sf20} 
        \put(65,9){\line(0,-1){2}}
        \put(67.5,9){\line(0,-1){1}}
        \put(52,1){{\Large\it l} }
        \put(11,9){\line(0,1){64}}
        \put(11,11){\line(-1,0){1}}
        \put(3,12){\small\sf300}
        \put(11,13){\line(-1,0){2}}
        \put(11,15){\line(-1,0){1}}
        \put(11,17){\line(-1,0){1}}
        \put(11,19){\line(-1,0){1}}
        \put(11,21){\line(-1,0){1}}
        \put(11,23){\line(-1,0){2}}
        \put(11,25){\line(-1,0){1}}
        \put(11,27){\line(-1,0){1}}
        \put(11,29){\line(-1,0){1}}
        \put(11,31){\line(-1,0){1}}
        \put(3,32){\small\sf200}
        \put(11,33){\line(-1,0){2}}
        \put(11,35){\line(-1,0){1}}
        \put(11,37){\line(-1,0){1}}
        \put(11,39){\line(-1,0){1}}
        \put(11,41){\line(-1,0){1}}
        \put(11,43){\line(-1,0){2}}
        \put(11,45){\line(-1,0){1}}
        \put(11,47){\line(-1,0){1}}
        \put(11,49){\line(-1,0){1}}
        \put(11,51){\line(-1,0){1}}
        \put(3,52){\small\sf100}
        \put(11,53){\line(-1,0){2}}
        \put(11,55){\line(-1,0){1}}
        \put(11,57){\line(-1,0){1}}
        \put(11,59){\line(-1,0){1}}
        \put(11,61){\line(-1,0){1}}
        \put(11,63){\line(-1,0){2}}
        \put(11,65){\line(-1,0){1}}
        \put(11,67){\line(-1,0){1}}
        \put(11,69){\line(-1,0){1}}
        \put(11,71){\line(-1,0){1}}
        \put(6.5,72){\small\sf0}
        \put(11,73){\line(-1,0){2}}
        \put(4,63){{\Large \it t} }
   \end{picture}
\end{center}
\caption{Probability $|\Phi _{jl}^{20}(t)|^2$ for $j=10$ as a surface diagram
in the $tl$ plane for $\Delta=0$, and 
$\Phi _{jl}^{20}(0)=\delta _{10,10}$. Height is indicated with a grey
scale such that black means zero height.}
\end{figure}

It is evident from this figure that
there is no coherent oscillation in the occupation numbers, the probability
is rapidly dispersed into the available Fock space. The behavior is
analogous to the one found in \cite{mol} for a ``spin wave'' in an optical
lattice, but there the dynamics was not followed very far (see also
discussion below for connection to the Heisenberg chain). Notice that the
time evolution is not translationally invariant at intermediate time scales,
and thus exhibits \cite{ir} ``aging''. There is however an almost complete recovery of
the initial probability distribution at a late time (which is repeated
periodically). The small wiggles in the probability that are clearly visible
at very early times are the ``Rabi oscillations'' of \cite{bt}.

 For non-zero effective detuning $\Delta \not =0$, Eq.~(\ref{phe}) can
 also be solved exactly. For the same initial condition as above, 
   \begin{equation}
      \Phi _{jl}^N(t) =
      \sum_{k=1}^{N+1}e^{-itE_k}\frac{B(\nu _k,\lambda ;j+1)
      B(\nu_k,\lambda ;l+1)}{\sum_{n=1}^{N+1}B^2(\nu _k,\lambda ;n)},
   \label{ehl}
   \end{equation}
 in which $\lambda =g/\Delta $, and $B(\nu ,\lambda ;k)$ are Lommel
 polynomials. In terms of the Bessel functions of the first and second kind
   $
      B(\nu ,\lambda ;k) =
      \frac{\pi \lambda }2 [J_\nu (\lambda )Y_{\nu +m}(\lambda) -
      J_{\nu +m}(\lambda )Y_\nu (\lambda )]
   $.
 The boundary conditions given
 below Eq.~(\ref{phe}) mean that $B(\nu ,\lambda ;N+2)=0$, and the $N+1$
 roots of this equation are the $\nu _k$. The spectrum $E_k$ is now 
      $ E_k=-\Delta (\nu _k+1)$. 
For small 
  $N/\lambda $ 
this spectrum is similar to the Bloch spectrum found above for zero effective detuning. 
There is a cross-over around 
  $N\approx \lambda $ 
into a linear spectrum: 
$E_k\approx \Delta k$ for $N/\lambda \gg 1$. 
This latter kind of spectrum is also called the Wannier-Stark ladder.

For $N/\lambda \ll 1$ we thus find the probability $|\Phi^N_{jl}(t)|^2$
behaves very similarly to that for zero effective detuning shown in Fig.~1.
Around $\lambda\approx N$ there is a transition to a completely different
behavior. An excitation composed of a superposition of states that are 
localized in the Fock space is formed in the system such that coherent
oscillation appears in the occupation probabilities of the two traps. 

\noindent
\unitlength=1mm
\begin{figure}[hbt]
\begin{center}
    \begin{picture}(68,71)
        \put(12,10){\psfig{file=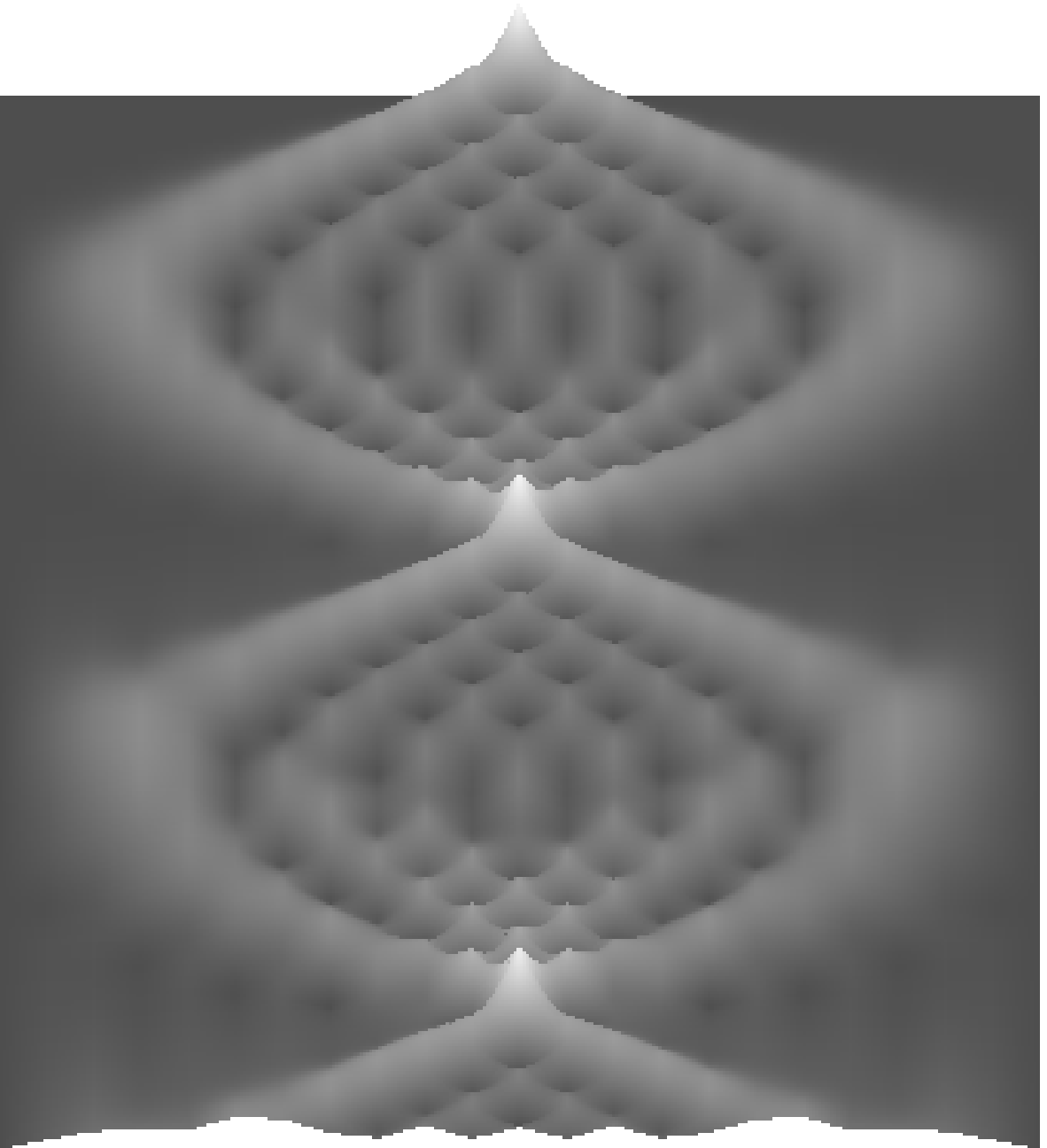,height=58mm,width=56mm}}
        \put(11,9){\line(1,0){59}}   
        \put(12.5,9){\line(0,-1){1}}
        \put(14.5,4){\small\sf0}
        \put(15,9){\line(0,-1){2}}
        \put(17.5,9){\line(0,-1){1}}
        \put(20,9){\line(0,-1){1}}
        \put(22.5,9){\line(0,-1){1}}
        \put(25,9){\line(0,-1){1}}
        \put(27.5,9){\line(0,-1){2}}
        \put(30,9){\line(0,-1){1}}
        \put(32.5,9){\line(0,-1){1}}
        \put(35,9){\line(0,-1){1}}
        \put(37.5,9){\line(0,-1){1}}
        \put(38.5,4){\small\sf10}
        \put(40,9){\line(0,-1){2}}
        \put(42.5,9){\line(0,-1){1}}
        \put(45,9){\line(0,-1){1}}
        \put(47.5,9){\line(0,-1){1}}
        \put(50,9){\line(0,-1){1}}
        \put(52.5,9){\line(0,-1){2}}
        \put(55,9){\line(0,-1){1}}
        \put(57.5,9){\line(0,-1){1}}
        \put(60,9){\line(0,-1){1}}
        \put(62.5,9){\line(0,-1){1}}
        \put(63.5,4){\small\sf20} 
        \put(65,9){\line(0,-1){2}}
        \put(67.5,9){\line(0,-1){1}}
        \put(52,1){{\Large\it l} }
        \put(11,9){\line(0,1){54}}
        \put(11,10.5){\line(-1,0){2}}
        \put(11,12){\line(-1,0){1}}
        \put(11,13.5){\line(-1,0){1}}
        \put(11,15){\line(-1,0){1}}
        \put(11,16.5){\line(-1,0){1}}
        \put(5,17){\small\sf60}
        \put(11,18){\line(-1,0){2}}
        \put(11,19.5){\line(-1,0){1}}
        \put(11,21){\line(-1,0){1}}
        \put(11,22.5){\line(-1,0){1}}
        \put(11,24){\line(-1,0){1}}
        \put(11,25.5){\line(-1,0){2}}
        \put(11,27){\line(-1,0){1}}
        \put(11,28.5){\line(-1,0){1}}
        \put(11,30){\line(-1,0){1}}
        \put(11,31.5){\line(-1,0){1}}
        \put(5,32){\small\sf40}
        \put(11,33){\line(-1,0){2}}
        \put(11,34.5){\line(-1,0){1}}
        \put(11,36){\line(-1,0){1}}
        \put(11,37.5){\line(-1,0){1}}
        \put(11,39){\line(-1,0){1}}
        \put(11,40.5){\line(-1,0){2}}
        \put(11,42){\line(-1,0){1}}
        \put(11,43.5){\line(-1,0){1}}
        \put(11,45){\line(-1,0){1}}
        \put(11,46.5){\line(-1,0){1}}
        \put(5,47){\small\sf20}
        \put(11,48){\line(-1,0){2}}
        \put(11,49.5){\line(-1,0){1}}
        \put(11,51){\line(-1,0){1}}
        \put(11,52.5){\line(-1,0){1}}
        \put(11,54){\line(-1,0){1}}
        \put(11,55.5){\line(-1,0){2}}
        \put(11,57){\line(-1,0){1}}
        \put(11,58.5){\line(-1,0){1}}
        \put(11,60){\line(-1,0){1}}
        \put(11,61.5){\line(-1,0){1}}
        \put(6.5,62){\small\sf0}
        \put(11,63){\line(-1,0){2}}
        \put(4,55){{\Large \it t} }
   \end{picture}
\end{center}
\caption{Probability $|\Phi _{jl}^{20}(t)|^2$ for $j=10$ as a surface
diagram in the $tl$ plane for $\lambda=5$, and
 $\Phi _{jl}^{20}(0)=\delta _{10,10}$. Height is indicated with a grey
scale such that black means zero height.}
\end{figure}

We show $|\Phi^N_{jl}(t)|^2$ for $N/\lambda=4$ in Fig.~2, where these features
of the solution are clearly visible. For increasing $N/\lambda$ the number of
localized states in the superposition decreases, and the
probability becomes more and more localized around the center line (i.e. the
initial value). This is evident from Fig.~3, which shows $|\Phi^N_{jl}(t)|^2$
for $N/\lambda=10$.

\noindent
\unitlength=1mm
\begin{figure}[hbt]
\begin{center}
    \begin{picture}(70,71)
        \put(12,10){\psfig{file=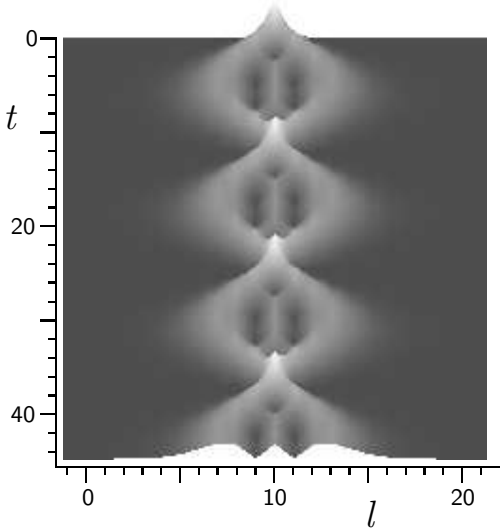,height=61mm,width=56mm}}
        \put(11,9){\line(1,0){59}}   
        \put(12.5,9){\line(0,-1){1}}
        \put(14.5,4){\small\sf0}
        \put(15,9){\line(0,-1){2}}
        \put(17.5,9){\line(0,-1){1}}
        \put(20,9){\line(0,-1){1}}
        \put(22.5,9){\line(0,-1){1}}
        \put(25,9){\line(0,-1){1}}
        \put(27.5,9){\line(0,-1){2}}
        \put(30,9){\line(0,-1){1}}
        \put(32.5,9){\line(0,-1){1}}
        \put(35,9){\line(0,-1){1}}
        \put(37.5,9){\line(0,-1){1}}
        \put(38.5,4){\small\sf10}
        \put(40,9){\line(0,-1){2}}
        \put(42.5,9){\line(0,-1){1}}
        \put(45,9){\line(0,-1){1}}
        \put(47.5,9){\line(0,-1){1}}
        \put(50,9){\line(0,-1){1}}
        \put(52.5,9){\line(0,-1){2}}
        \put(55,9){\line(0,-1){1}}
        \put(57.5,9){\line(0,-1){1}}
        \put(60,9){\line(0,-1){1}}
        \put(62.5,9){\line(0,-1){1}}
        \put(63.5,4){\small\sf20} 
        \put(65,9){\line(0,-1){2}}
        \put(67.5,9){\line(0,-1){1}}
        \put(52,1){{\Large\it l} }
        \put(11,9){\line(0,1){57}}
        \put(11,11){\line(-1,0){1}}
        \put(11,13.5){\line(-1,0){1}}
        \put(5,15){\small\sf40}
        \put(11,16){\line(-1,0){2}}
        \put(11,18.5){\line(-1,0){1}}
        \put(11,21){\line(-1,0){1}}
        \put(11,23.5){\line(-1,0){1}}
        \put(11,26){\line(-1,0){1}}
        \put(11,28.5){\line(-1,0){2}}
        \put(11,31){\line(-1,0){1}}
        \put(11,33.5){\line(-1,0){1}}
        \put(11,36){\line(-1,0){1}}
        \put(11,38.5){\line(-1,0){1}}
        \put(5,40){\small\sf20}
        \put(11,41){\line(-1,0){2}}
        \put(11,43.5){\line(-1,0){1}}
        \put(11,46){\line(-1,0){1}}
        \put(11,48.5){\line(-1,0){1}}
        \put(11,51){\line(-1,0){1}}
        \put(11,53.5){\line(-1,0){2}}
        \put(11,56){\line(-1,0){1}}
        \put(11,58.5){\line(-1,0){1}}
        \put(11,61){\line(-1,0){1}}
        \put(11,63.5){\line(-1,0){1}}
        \put(7,65){\small\sf0}
        \put(11,66){\line(-1,0){2}}
        \put(4,54){{\Large\it t} }
   \end{picture}
\end{center}
\caption{Probability $|\Phi _{jl}^{20}(t)|^2$ for $j=10$ as a surface
diagram in the $tl$ plane for $\lambda=2$, and
 $\Phi _{jl}^{20}(0)=\delta _{10,10}$. Height is indicated with a grey
scale such that black means zero height.}
\end{figure}

We could call the localized oscillatory solution as a ``trapped
quantum soliton''. We believe that this solution should be experimentally
observable in coupled traps with ultracold atoms, in which the parameter
$\lambda$ can fairly easily be controlled by controlling either $\Delta$
or $g$ or both. Notice also that the states that form the
superposition are entangled. This system would thus be suitable for analyzing 
entanglement, and perhaps even quantum computing as ``engineering'' of states 
is possible.

It is also instructive to consider the limit in which one of the coupled
 traps becomes macroscopic (classical): $N\to \infty $ (i.e. either
$N_1$ or $N_2$). For simplicity we consider this case only in the limit 
of weakly interacting particles and set $c=0$.
The correlator 
      $ \Phi _{jl}^\infty (t)= \langle j|\exp (-itH_\infty) |l \rangle $
 satisfies the same equation Eq.~(\ref{phe}) as for finite $ N $
 but now the boundary conditions are
      $\Phi_{jl}^\infty (t)=0$ for $j,l=-1$.
 The Hamiltonian becomes in this limit 
   \begin{equation}
      \hat H_\infty =
      \Delta \hat N-\frac g2(\phi +\phi ^{\dagger }).
   \label{eph2}
   \end{equation}
 Notice that the (new) number operator $\hat N$ does not any more commute
 with the Hamiltonian (this is similar to the corresponding limit in the
 Bogoliubov theory \cite{b} of Bose condensates). If we express this Hamiltonian
 Eq.~(\ref{eph2}) in the Fock space, we find that 
   \begin{equation}
      \hat H_\infty =
      \Delta \sum n|n\rangle \langle n| -
      \frac g2\sum 
      \left( |n\rangle \langle n+1|+|n+1\rangle \langle n|\right) .
   \label{eph3}
   \end{equation}
This form of the Hamiltonian appears in many problems related to trapped
gases of ultracold atoms, e.g., in the tight binding limit of a ring-shaped
trap with a moving defect  \cite{buch}.

The correlator $\Phi _{jl}^\infty (t)$ can again be solved exactly with the
 result 
   \begin{equation}
      \Phi _{jl}^\infty (t) =
      \sum_{k=1}^\infty e^{-itE_k}
      \frac{J_{j+1+\nu_k}(\lambda )J_{l+1+\nu _k}(\lambda )}
      {\sum_{n=1}^\infty J_{n+\nu_k}^2(\lambda )} ,
   \label{ninfh}
   \end{equation}
 where summation is over the $\nu _k$ determined by the solution of the
``index equation'' $J_{\nu _k}(\lambda )=0$. The spectrum is now 
      $ E_k+\Delta =-\Delta |\nu _k| $,
 which is very similar to the Wannier-Stark spectrum for small $k$.
 The solution is also in this case a coherent oscillation in the
 occupation number of the finite trap similarly to the solution shown in
 Fig.~3. The same ``quantum features'' of the solution can thus be realized
 with only one trap with a small number of weakly interacting particles.

For zero effective detuning coherent oscillations disappear as expected, and
we find that asymptotically $|\Phi^{\infty}_{jl}(t)|$ decays as $t^{-3/2}$.
The dynamical exponent $z=3/2$ also appears in nonequilibrium growth 
processes with one-dimensional interfaces \cite{sp}.

We shall finally make a connection to quantum computing even though a
detailed analysis of this aspect of the problem cannot be given here. We
show here that the correlators $\Phi^N_{jl}(t)$ considered above can be
mapped exactly to the corresponding correlators of an isotropic Heisenberg
spin $1/2$ chain in a linear field, for which all logical operations of quantum
computing have already been \cite{vin} constructed.

Consider the one spin-flip correlator (related to the ``{\small NOT}'' operation) 
   \begin{equation}
      F_{jl}^{N+1}(t)\equiv
      \langle 0\mid \sigma _j^{+}e^{-it\hat H_{xxx}}\sigma
     _l^{-}\mid 0\rangle ,  \label{ecor}
   \end{equation}
 where the ground state $ |0 \rangle $ is that with all spins pointing up,
 and the Heisenberg Hamiltonian $H_{xxx}$ is given by 
   \begin{eqnarray}
      &&\hat H_{xxx} =
      -\frac g2 \left[ \sum_{n=1}^N\sigma _n^{+}\sigma _{n+1}^{-}+ 
      \sigma_n^{-}\sigma _{n+1}^{+}+
      \frac 12(\sigma _n^z\sigma _{n+1}^z-1) \right.
   \nonumber
   \\
      &&\left.
      +\frac 12 (\sigma _1^z+\sigma _{N+1}^z-2) \vphantom{\sum_{n=1}^N}
      \right]
      - \frac12 \sum_{n=1}^{N+1}h_n(\sigma _n^z-1).
   \label{eXXX}
   \end{eqnarray}
Here $\sigma _n^{\pm },\sigma _n^z$ are Pauli spin matrices, and $h_n$ is an
external magnetic field. The constant factors in the Hamiltonian (\ref {eXXX}) 
are added to ensure that $H_{xxx}|0 \rangle =0$. As before we can derive an
equation of motion for correlator (\ref{ecor}), and find for fixed index $j$
that
\begin{eqnarray}
      -i\frac 2g\frac d{dt}F_{jl}^{N+1}(t) =
      F_{jl+1}^{N+1}(t)+F_{jl-1}^{N+1}(t)
   \nonumber\\
   -2\left( 1+\frac{h_l} g\right) F_{jl}^{N+1}(t)
   \label{eseq}
   \end{eqnarray} 
with a similar equation for fixed $l$. The initial condition to be applied
here is $F_{jl}^{N+1}(0)=\delta _{jl}$, and the boundary conditions
appropriate for the lattice of $N+1$ sites are $F_{jl}^{N+1}(t)=0;j,l=0,N+2$%
. As the constant factor in the last term on the right side of Eq.~(\ref
{eseq}) only gives a constant global phase in the solution, it is evident
that the absolute value of a solution of this equation for a constant
magnetic field is identical with the corresponding solution of
Eq.~(\ref{phe}) with $\Delta=0$, and with that of Eq.~(\ref{phe}) 
with $\Delta\not=0$ for a linear magnetic field $h_n=nh$. There is
obviously a
one-to-one mapping from the Heisenberg dynamics into that of coupled traps
with ultracold atoms. The number of lattice sites in the Heisenberg chain
plays the role of total particle number in the traps. Thus, e.g., the limit
$ N\to \infty $ in the coupled traps, with Hamiltonian (\ref{eph2}), 
corresponds to the limit of semi-infinite chain in the Heisenberg problem.
Notice that for a non-zero linear magnetic field, the localized coherent
oscillations of Figs.~2 and 3 appear also in the Heisenberg chain. 

In conclusion, we have considered a model for two traps with ultracold atoms
and with a strong Josephson coupling between them. We find that for
non-zero effective detuning in the trap energies there is a transition to
localized coherent oscillation in the occupation probabilities of the traps.
Below the transition, i.e. for $N/\lambda \ll 1$, the dynamics is
dispersive as
in the case of vanishing offset. We also found that there is a one-to-one
mapping between these dynamics and those of an isotropic Heisenberg chain in
an appropriate magnetic field. This opens up the possibility in principle of
realizing quantum computation, or simulation, in coupled-trap systems with
only the detuning and the coupling of the traps as the
relevant parameters.

We thank A.G. Pronko and one of us, M.B.Z., V.~Cheianov for fruitful 
discussions. N.M.B. would like to thank the University of Jyv\"askyl\"a,
and M.B.Z. NORDITA, for support and hospitality.
This work was partially supported by the Russian
Foundation for Basic Research (Project No. 01-01-01045), and the
Academy of Finland (Project No. 44875).

\end{document}